\documentclass[journal,10pt]{IEEEtran}
\usepackage{tabularx}
\usepackage[noadjust]{cite}
\usepackage{tabularx}
\usepackage{diagbox}
\usepackage{multirow}
\usepackage{booktabs}
\usepackage{hhline}
{}
{}
{}
\usepackage{setspace}
\usepackage{amsthm,amsmath,amssymb,lipsum}
\usepackage{mathrsfs}
\usepackage{stfloats}
\usepackage{enumerate}
\usepackage[ruled,vlined,linesnumbered]{algorithm2e}  
\usepackage{bm}   
\newcommand{\RNum}[1]{\uppercase\expandafter{\romannumeral #1\relax}}
\usepackage{booktabs}
\usepackage{xcolor}
\usepackage{balance}
\definecolor{flexred}{rgb}{0, 0, 0}

\ifCLASSINFOpdf
  \usepackage[pdftex]{graphicx}
  \graphicspath{{./figure/}}
  \DeclareGraphicsExtensions{.pdf,.jpeg,.png}
\else
\fi

\ifCLASSOPTIONcompsoc
\usepackage[caption=false,font=normalsize,labelfont=sf,textfont=sf]{subfig}
\else
\usepackage[caption=false,font=footnotesize]{subfig}
\fi

\usepackage{mathtools,amssymb}

\usepackage{cuted}
\definecolor{flexred}{rgb}{0.9, 0, 0}

\makeatletter
\newcommand{\biggg}{\bBigg@{1.2}}

\makeatother

\usepackage{soul}
\begin{document}


\title{Throughput Maximization for Intelligent Refracting Surface Assisted mmWave High-Speed Train Communications}

\author{
Jing~Li,~
Yong~Niu,~\IEEEmembership{Senior Member,~IEEE,}
Hao~Wu,~\IEEEmembership{Member,~IEEE,}
Bo~Ai,~\IEEEmembership{Fellow,~IEEE,}
Ruisi~He,~\IEEEmembership{Senior Member,~IEEE,}
Ning~Wang,~\IEEEmembership{Member,~IEEE,}
Sheng~Chen,~\IEEEmembership{Life Fellow,~IEEE}
\thanks{This work was supported
in part by the National Key Research and Development Program of China under Grant 2021YFB2900301, in part by the National Key Research and Development Program of China under Grant 2020YFB1806903, in part by the National Natural Science Foundation of China under Grants 62221001, 62231009, and U21A20445, in part by the Fundamental Research Funds for the Central Universities, China, under Grants 2022JBQY004 and 2022JBXT001, in part by the Fundamental Research Funds for the Central Universities under Grants 2023JBMC030, 2022YJS114, and in part by the Science and Technology Research and Development Plan of China Railway Co., Ltd under Grant K2022G018.  \emph{(Corresponding Author: Yong Niu.)}}
\thanks{Jing Li, Yong Niu, Hao Wu, Bo Ai and Ruisi He are with the State Key Laboratory of Advanced Rail Autonomous Operation, Beijing Jiaotong University, Beijing 100044, China, and also with Beijing Engineering Research Center of High-speed Railway Broadband Mobile Communications, Beijing 100044, China (E-mails: jinglee@bjtu.edu.cn, niuy11@163.com, hwu@bjtu.edu.cn, boai@bjtu.edu.cn, ruisi.he@bjtu.edu.cn).} %
\thanks{Ning Wang is with School of Information Engineering, Zhengzhou University, Zhengzhou 450001, China (E-mail: ienwang@zzu.edu.cn).} %
\thanks{Sheng Chen is with School of Electronics and Computer Science, University of Southampton, Southampton SO17 1BJ, UK (E-mail: sqc@ecs.soton.ac.uk).} %
\vspace*{-5mm}
}

\maketitle

\begin{abstract}
With the increasing demands from passengers for data-intensive services, millimeter-wave (mmWave) communication is considered as an effective technique to release the transmission pressure on high speed train (HST) networks. However, mmWave signals encounter severe losses when passing through the carriage, which decreases the quality of services on board. In this paper, we investigate an intelligent refracting surface (IRS)-assisted HST communication system. Herein, an IRS is deployed on the train window to dynamically reconfigure the propagation environment, and a hybrid time division multiple access-nonorthogonal multiple access scheme is leveraged for interference mitigation. We aim to maximize the overall throughput while taking into account the constraints imposed by base station beamforming, IRS discrete phase shifts and transmit power. To obtain a practical solution, we employ an alternating optimization method and propose a two-stage algorithm. In the first stage, the successive convex approximation method and branch and bound algorithm are leveraged for IRS phase shift design. In the second stage, the Lagrangian multiplier method is utilized for power allocation. Simulation results demonstrate the benefits of IRS adoption and power allocation for throughput improvement in mmWave HST networks.
\end{abstract}


\IEEEpeerreviewmaketitle

\section{Introduction}\label{S1}

As an efficient, green and eco-friendly transport mode, high-speed train (HST) has attracted worldwide attention in the past two decades. By 2022, the total millage of HST rail lines, with the speed of train exceeding 200 km/h, has reached 56,000 km and is expected to double in 30 years \cite{CNN}. In many countries, like China and Spain, HSTs link the major city clusters, fasten long-distance travel, and are changing people's mobility habits deeply. Fast growing number of passengers has demanded much stricter safe operation as well as multimedia services from railway systems, which powers the sustainable development of wireless communications for HST systems \cite{H2016}. Initially, the Global System for Mobile Communications Railway \cite{WCX} provisioned services of voice communications and control signaling transmission but failed to satisfy personal broadband applications. To handle this problem, Long Term Evolution for Railway \cite{WCX} was introduced at the beginning of 2010s, which supports megabits-per-second order data rate, but still does not meet the requirements of massive connectivity and intensive data exchange for future HST systems \cite{WCX, H2016}. The fifth-generation (5G) technologies are expected to ensure real-time and high-quality transmissions for both passengers and railway operation messages, to start a new era in HST communications \cite{AB2015}.

Exploiting millimeter-wave (mmWave) communications, 5G for Railway \cite{HRS2022} has the potential to offer multimedia services that require multi-gigabit rate, e.g., video conference, live broadcast, online gaming, etc., thereby enhancing the customer experience. In addition, mmWave integrated with multiple-input multiple-output (MIMO) and beamforming techniques is capable of offering comprehensive perception, interconnection, and information interaction among railway users and infrastructures, which subsequently contributes to the construction of intelligent transportation system (ITS) \cite{HRS2022}. However, given the propagation characteristics of mmWave signals and dynamic nature of railway scenarios, there are many challenges to overcome in order to maintain high-quality on-board services. Specifically, experiencing severe path loss, the coverage radius of mmWave is limited, and frequent handover occurs in train-to-ground (T2G) communications. Moreover, mmWave signals suffer from serious penetration losses when passing through solid materials, including glasses, metals and trees, and consequently they are vulnerable to blockage \cite{RW2013}. During mobility, the quality of service (QoS) degrades significantly once line-of-sight (LoS) links between passengers and the base station (BS) are blocked by the carriage or other obstacles \cite{ZJY}. To this end, many researchers deploy mobile relay stations (MRSs) on the train to combat  penetration losses and control handover, whereas MIMO technologies are typically involved as well \cite{WCX, Song, AB2014}. These solutions, however, require extra hardware cost and energy consumption, while introducing transmission delay.

Fortunately, the emerging technology of reconfigurable intelligent surface (RIS) is envisioned to smartly control the mmWave propagation environment to improve the communication quality. Therefore, there have been many studies \cite{LPP, a2, XJP} on RIS-aided HST networks in recent years, most of which considered fixing RIS panels on the railroad side. However, this deployment limits the service time for the RIS to assist fast-moving users, thus constraining its beneficial effects \cite{WGP2023}. 
To address this issue, the authors of \cite{ICCT2022} deployed a RIS on the HST to follow the in-train user and exploited its reflective properties to mitigate the delay spread, which still fails to resolve the challenging problem of signal attenuation caused by train carriages. 
Inspired by \cite{WGP2023, Slow}, we investigate the feasibility of employing intelligent refracting surfaces (IRSs) in future railway communications. That is, dynamically reconfiguring the HST channel by adjusting the coefficient of each refractive element, which promises to enable robust wireless links and improve the QoS (e.g., throughput, reliability, delay, etc. \cite{Q1,Q2,Q3}) in the train without extra power consumption \cite{transparent2}. For this reason, we study to enhance the throughput of downlink HST communications by a window-deployed IRS that refracts incident signals to users during mobility, with the network throughput being the QoS metric.

In addition, non-orthogonal multiple access (NOMA) as a key technology in 5G has attracted significant attention. Due to its ability to improve the capacity and spectral efficiency of wireless communication systems, there have been several studies employing the NOMA scheme in HST networks. For instance, the authors of \cite{VTC2023} investigated the outage performance of NOMA HST systems and the work \cite{MT2022} applied NOMA technology to enhance the uplink energy efficiency of HST communications. 
Nevertheless, how to integrate NOMA technique in the dynamic IRS-HST scenario is still a current issue. In this case, a hybrid time division multiple access (TDMA)-NOMA scheme \cite{TDMA-NOMA} is leveraged in our model, where the available time is divided into non-overlapping frames and multiple users in each cluster access their spectrum using NOMA in each frame.
Such a hybrid transmission scheme effectively eliminates transmission interference and fully exploits the advantages of IRS in HST communications, 
because: 1) in each frame, the mutual interference (MUI) between users is reduced and power domain multiplexing is realized, thereby improving the access capacity \cite{YZhu} while satisfying different performance requirements; 2) it enables the frame-by-frame updating of beamforming and phase shift design, which fits well with the dynamic environment.
Also a power control strategy is proposed to further improve the system throughput and to meet the requirements of power-efficient, green railway construction.  
The main contributions of this paper are summarized as follows.
 
\begin{itemize}
\item We study an IRS-assisted hybrid  TDMA-NOMA mmWave HST communication system, where a transparent IRS is deployed to assist T2G communications. Based on the practical position-prediction model and channel model, we formulate the throughput maximization problem that takes into account constraints on BS beamforming, IRS refraction coefficients and transmit power.
\item We propose a two-stage optimization algorithm to solve this problem. Specifically, first we alternately optimize beamforming and IRS phase shifts in each frame, where the successive convex approximation (SCA) and branch and bound (BB) algorithm are utilized. Based on these optimization results and incorporated with the characteristics of high-speed mobility, we perform local power allocation in every fixed interval, to improve the overall throughput effectively.
\item By comparing the achievable performance of the proposed algorithm with other existing schemes given different numbers of users, IRS sizes, Rician K-factors and quantization bits, the simulation results reveal the effectiveness of IRS adoption and power allocation for throughput improvement in the HST scenario.
\end{itemize}
 
The remainder of this paper is organized as follows. Section~II summarizes the related work. Section~\ref{S3} introduces the system model and then formulates the throughput maximization problem. In Section~\ref{S4}, a two-stage solution is proposed based on alternating optimization and Lagrangian multiplier method. Simulation results and discussions are presented in Section~\ref{S5}. Finally, Section~\ref{S6} concludes the paper. 

\emph{Notations:} We use upper-case and lower-case bold face letters to indicate matrices and vectors/vector sets; $(\cdot)^{\rm T}$ and $(\cdot)^{\rm H}$ represent transpose and Hermitian transpose, respectively.

\section{Related Work}\label{S2}
As a new type of RIS, the emerging IRS offers a new paradigm for wireless communications. Typically, each IRS is composed of two layers with a metallic backplane in the middle, which transfers the incident energy from one side to the other side while shielding the signal reflection \cite{Related1}. By controlling all refractive elements, the IRS serves users located on the opposite side of the Tx with respect to the surface, which is different from conventional RISs working in reflective mode. 
Also leveraging its refractive characteristic is promising to overcome the limits in existing wireless communications, regarding achievable indoor data rate, energy efficiency enhancement, and the coverage expansion.  
Therefore, growing awareness has been paid on the IRS design and application in recent years. 
Initially, efficient refractive metasurfaces were designed based on physical property requirements in \cite{2016ITS, PC}, paving the way for IRS solutions. 
Then in 2020, NTT DOCOMO verfied the prototype transparent dynamic metasurface operating at 28 GHz, contributing to the construction and optimization of 5G network \cite{NTT}.  
Furthermore, the authors of \cite{transparent2} implemented optically transparent metasurfaces to expand the mmWave coverage, which can be attached to existing walls and glass windows.
 
Moreover, IRS is capable of aiding mobile communications in vehicular networks. For example, the work \cite{Slow} successfully deployed the IRS on a moving vehicle for enhancing the transmission rate and reliability between in-vehicle users and the roadside BS by channel estimation and refraction optimization. A few researchers have also noted IRS-assisted HST communications. The study \cite{ZJY} investigated the possibility of IRS technology to address the unique design challenges of HST networks and validated its benefits. Likewise, the authors of \cite{WGP2023} presented a low-complexity channel estimation method with Doppler shifts recovery, which is critical for timely reconfiguration of the wireless environment through IRS coefficient adjustment. These efforts of applying IRSs however are still limited to channel design and seldom involve signal processing or resource management techniques for performance improvement.
In this case, we consider beamforming optimization and resource allocation for IRS-assisted HST communications, with the system design, transmission mode and objective function different from earlier attempts \cite{MZ, XJP, AR, GML, LT} using the RIS reflective modes. To improve the obtained throughput, we also creatively combine the idea of alternating optimization with the SCA method and Lagrangian multiplier method, shown to be effective in such a high-mobility scenario.

\begin{figure}[!b]
\vspace*{-3mm}
\centering
\includegraphics[width=0.95\columnwidth]{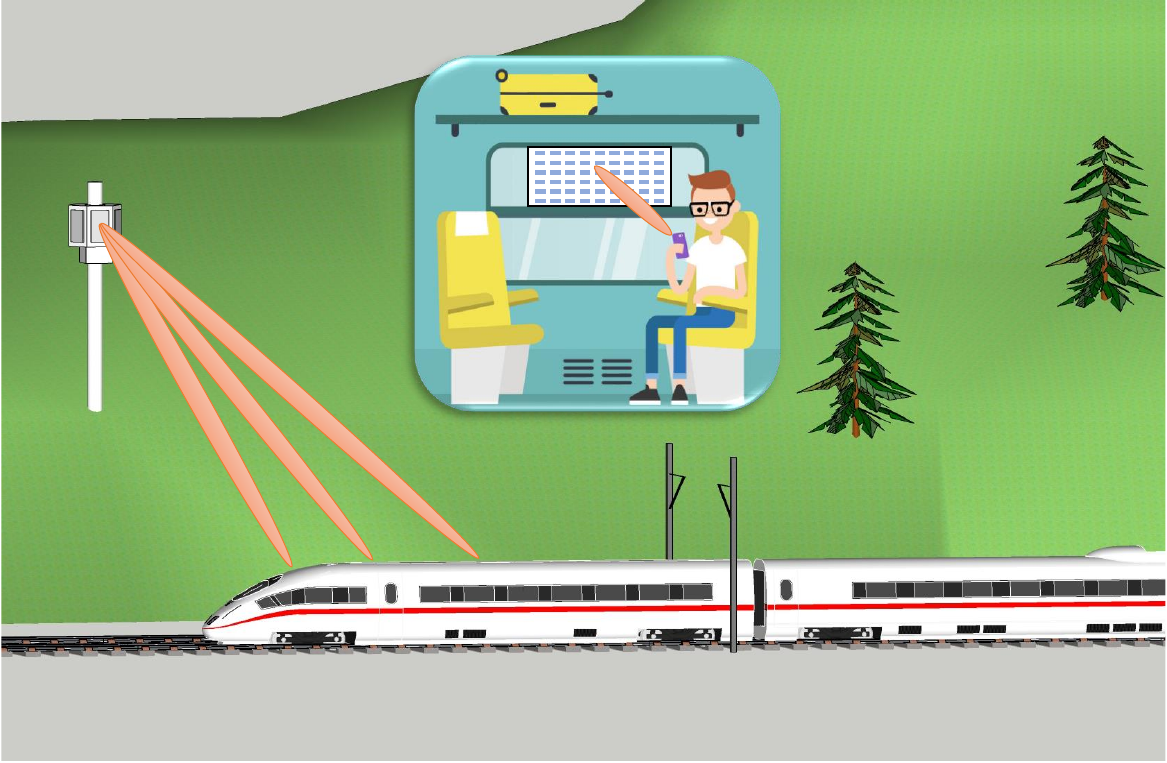}
\vspace*{-1mm}
\caption{IRS-assisted mmWave HST communication system.}
\label{fig:system} 
\vspace*{-1mm}
\end{figure}

\section{System Model}\label{S3}

We investigate the IRS-assisted mmWave HST communication system illustrated in Fig.~\ref{fig:system}, where $I$ users on board are served by a track-side BS with $L$ antennas. To mitigate penetration losses, an IRS with $M$ refractive elements is mounted on the train window to aid downlink power transfer and communications. Since practical approaches on channel state information (CSI) acquisition for RIS-aided MIMO systems have been proposed \cite{LPP, BYTVT}, it is reasonable to suppose that the CSI of all channels involved is perfectly known at the BS and can be transmitted to the IRS controller over a dedicated control channel. The controller independently adjusts the biasing voltage of each element based on the CSI so as to refract incident signals to desired directions, thus improving the network throughput on board.
For ease of exposition, we assume that the train runs at a constant speed of $v$ and passes through the cell with a coverage radius of $R$ in time $t=2R/v$. In practice, users remain relative stationary with each other in a frame while the BS transmits data streams concurrently.

In our system, a hybrid TDMA-NOMA transmission scheme is leveraged to balance the complexity and performance. 
Without loss of generality, we adopt the uniform distribution model of traffic load \cite{WCNC}, where users' access requests are uniformly distributed in each frame and proposed only once. Thus, $I$ users can be equally grouped into $K$ clusters based on their requesting time, and $N$ users in each cluster simultaneously access the network by employing NOMA.
Denoting the duration of each TDMA frame as $\tau$, different clusters communicate with the BS in an orthogonal TDMA manner, which decreases MUI and contributes to the performance improvement. The communication distance and channel model are updated frame by frame, which are specified as follows. 

\subsection{Distance-Prediction Model}\label{S2.1}
\begin{figure}[!htp]
\centering
\includegraphics[width=0.95\columnwidth]{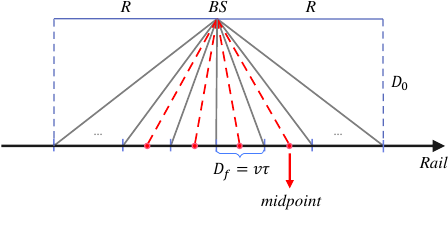}
\vspace*{-5mm}
\caption{The mobility trajectory of HST.}
\label{fig:Distance} 
\vspace*{-1mm}
\end{figure}
Because of the short frame duration, it is reasonable to assume that the communication distance between the BS and the train remains approximately constant in a frame, but it changes from frame to frame. Therefore, we consider the mobility model illustrated in Fig.~\ref{fig:Distance}, where the distance from the BS to the rail is denoted by $D_0$, and the position of the train in frame $k$ can be approximated with the midpoint of the corresponding segment. Thus, we estimate the distance between the BS and IRS in frame $k$, denoted as $D_k$, by
\begin{equation}\label{equ:d(k)} 
D_k = \left\{ \begin{array}{cl}
\left(\left(R-(k-\frac{1}{2})D_{f} \right)^2 +D^2_0\right)^{\frac{1}{2}}, & k\leq \lfloor \frac{K}{2} \rfloor, \\
\left(\left((k-\frac{1}{2})D_{f}-R\right)^2 +D^2_0\right)^{\frac{1}{2}} , & k>\lfloor \frac{K}{2} \rfloor, \end{array} \right.
\end{equation}
with $D_f=v\tau$ denoting the distance advanced in a frame.

\subsection{Channel Model}\label{S2.2}

Consider a quasi-static fast fading channel, with the CSI remaining constant within a frame but changing in subsequent frames \cite{HuaMeng}. The channel gains from the BS to the IRS and from the IRS to users are modeled as Rician fading, where the contributions of LoS and non-LoS (NLoS) multipath components are taken into account \cite{Omni}. For the BS-IRS channel, Doppler effect is induced by the high-speed relative mobility between the fixed BS and the IRS mounted on the train, with which its channel gain matrix in frame $k$, $\textbf{G}_k\in\mathbb{C}^{M\times L}$ can be expressed as
\begin{equation}\label{equ:inter}  
\textbf{G}_k = \sqrt{\frac{K_f}{K_f+1}}\overline{\textbf{G}}_k+\sqrt{\frac{1}{K_f+1}}\widehat{\textbf{G}}_k,
\end{equation} 
where $K_f$ denotes the Rician K-factor. In (\ref{equ:inter}), $\overline{\textbf{G}}_k\in\mathbb{C}^{M\times L}$ represents the LoS component given by \cite{XJP}

\begin{align}
\setlength{\arraycolsep}{0.7pt}
\begin{array}{lc}
    \overline{\textbf{G}}_k=\vspace{0.5ex}
     \sqrt{h_0 D^{-\beta_1}_{k}}\\
     \begin{bmatrix}
 \alpha_k\big(\phi_{1,1}^{A},\omega_{1,1}^{A}\big)e^{j2\pi \zeta^k_{1,1}\tau} & \dots & \alpha_k\big(\phi_{1,L}^{A},\omega_{1,L}^{A}\big)e^{j2\pi \zeta^k_{1,L}\tau} \\
 \vdots & \ddots & \vdots \\
 \alpha_k\big(\phi_{M,1}^{A},\omega_{M,1}^{A}\big)e^{j2\pi \zeta^k_{M,1}\tau} & \dots & \alpha_k\big(\phi_{M,L}^{A},\omega_{M,L}^{A}\big)e^{j2\pi \zeta^k_{M,L}\tau}
\end{bmatrix},
 \end{array}
    \label{equ:inter1}
\end{align}
\\
where $h_0$ is the reference path loss at distance of 1\,m, $D_k$ is the distance between the BS and IRS, kept constant in frame $k$, and $\beta_1$ is the path loss exponent. $\alpha_k(\cdot ,\cdot )$ denotes the array response, $\phi^{A}_{m,l}$ and $\omega^{A}_{m,l}$ are the azimuth and elevation angles-of-arrival (AoAs), respectively, from the $l$-th BS antenna to the $m$-th IRS element at frame $k$, while $\zeta^k_{m,l}$ represents the Doppler shift corresponding to this path. In (\ref{equ:inter1}), the amplitude gain at IRS is neglected since signals also experience penetration losses when passing through the window.

Moreover, $\widehat{\textbf{G}}_k\in\mathbb{C}^{M\times L}$ represents the NLoS component matrix, whose elements follow the complex normal distribution with zero mean and unit variance \cite{NLOSNornmal}.

Similarly, the LoS component of the channel gain vector $\textbf{v}_{k,i}^{\rm H}\in \mathbb{C}^{1\times M}$ from the IRS to user $i$ in frame $k$ is given by
\begin{equation}\label{equ:inter2} 
\overline{\textbf{v}}_{k,i}^{\rm H}\! =\! \sqrt{h_0 d^{-\beta_2}_{k,i}}\! \begin{bmatrix}\alpha_{k,i}\big(\phi_{1,1}^{D},\omega_{1,1}^{D}\big), \,\dots \,, \alpha_{k,i}\big(\phi_{1,M}^{D},\omega_{1,M}^{D}\big)\end{bmatrix},
\end{equation}
where $d_{k,i}$ is the distance between the IRS and the $i$-th user, $\beta_2$ denotes the path loss exponent inside the carriage, $\phi_{1,m}^{D}$ and $\omega_{1,m}^{D}$ are the azimuth and elevation angles of departure (AoDs), respectively, from the $m$-th IRS element to the $i$-th user. The elements of the NLoS component vector $\widehat{\textbf{v}}_{k,i}^{\rm H}\in \mathbb{C}^{1\times M}$ also follow the complex normal distribution with zero mean and unit variance.

Given that authors in \cite{CEIRS} have proposed practical channel
estimation schemes compensating for Doppler shift in IRS-assisted mmWave system, we assume the Doppler effect can be eliminated. Also we suppose the perfect CSI for BS-IRS and IRS-user links are available, which is consistent to \cite{MZ}. Since refracting elements have no digital processing ability, we perform digital beamforming at the BS and analog beamforming at the IRS. Denote the transmit signal for user $j$, BS beamforming vector and IRS phase shift matrix in frame $k$ by $s_{k,j}$, $\textbf{f}_k\in\mathbb{C}^{L\times 1}$ and $\bm{\Theta}_k=\mathrm{diag}\left(e^{\textsf{j}\psi_{k,1}},\dots ,e^{\textsf{j}\psi_{k,M}}\right)$.
The received signal for user $i$ can be expressed as
\begin{align}\label{equ:V2VA} 
  r_{k,i} =& \textbf{v}^{\rm{H}}_{k,i} \bm{\Theta}_k \textbf{G}_{k} \textbf{f}_k \sum\limits_{j=1}^N \sqrt{P_k} s_{k,j} + w_i 
  = \underbrace{\textbf{v}^{\rm{H}}_{k,i} \bm{\Theta}_k \textbf{G}_{k} \textbf{f}_k \sqrt{P_k} s_{k,i}}_{\text{desired signal}} \nonumber \\ 
  & + \underbrace{\sum\limits_{j=1, j\neq i}^{N} \textbf{v}^{\rm{H}}_{k,i} \bm{\Theta}_k \textbf{G}_{k} \textbf{f}_k \sqrt{P_k} s_{k,j}}_{\text{user interference}} + w_i ,
\end{align}
where $P_k$ is the transmit power in frame $k$, and $w_i$ is the additive white Gaussian noise (AWGN) at user $i$, with zero mean and variance $\sigma^2$, and $E\big[|s_{k,j}|^2\big]=1$, $\forall k,j$. Besides, since the cluster-based NOMA \cite{CNOMA} is leveraged in this work, it is reasonable to assume that $N$ users in each cluster/frame are served by a common BS beamforming $\textbf{f}_k\in\mathbb{C}^{L\times 1}$.

Exploiting successive interference cancellation (SIC), the signal-to-interference-plus-noise ratio (SINR) received by user $i$ in frame $k$ can be written as
\begin{equation}\label{equ:V2VA1} 
\gamma_{k,i} = \frac{P_k\left|\textbf{v}^{\rm{H}}_{k,i} \bm{\Theta}_k \textbf{G}_k \textbf{f}_k\right|^2}{\sum\limits_{j=i+1}^{N} P_k\left|\textbf{v}^{\rm{H}}_{k,j} \bm{\Theta}_k \textbf{G}_{k} \textbf{f}_k\right|^2 + \sigma^2}.
\end{equation}
The aggregate achievable data rate in frame $k$ is given by \cite{Wu_etal2018}
\begin{align}\label{equ:R}  
R_k =& \sum\limits_{i=1}^{N} \log_2\left(1 + \gamma_{k,i} \right) \nonumber \\
=& \log_2\left(1 + \sum\limits_{i=1}^{N} \frac{P_k\left|\textbf{v}^{\rm{H}}_{k,i} \bm{\Theta}_k \textbf{G}_k \textbf{f}_k\right|^2}{\sigma^2}\right) .
\end{align}
\emph{Proof:} See Appendix A.
\subsection{Problem Formulation}\label{S2.3}

First we recap the system configuration and parameters. The BS has $L$ antennas, and the IRS has $M$ refractive elements. The system supports $I$ users, and they are divided equally into $K$ clusters, each having $N$ users, i.e., $I = K N$. The time duration $t$ is divided into $K$ TDMA frames, each having the duration of $\tau$, i.e., $t = K \tau$. Each group is assigned with a distinct frame, and the $N$ users of the same group share the same frame. Furthermore, the total transmit power over time duration $t$ is set to $P_{\rm sum}$. Each refractive element $\theta = e^{\textsf{j}\psi}$ of the IRS can take the value from the set of $2^e$ values, namely, the phase shift $\psi = \frac{2\pi a}{2^e-1}$, $a\in \{0,1,\dots ,2^e-1\}$, with the magnitude $\|\theta\|=1$, where $e$ denotes the quantization bits. We also define the transmit power vector over the $K$ TDMA frames by $\textbf{p}=\big[P_1 , \dots , P_K\big]^{\rm T}$.

Our goal is to maximize the overall throughput in time duration $t$ by jointly optimizing the beamforming, IRS phase shifts and power allocation. The problem is defined as
\begin{align} 
	& \mathcal{P}: \max\limits_{\textbf{p},\{\bm{\Theta}_k\}_{k=1}^K,\{\textbf{f}_k\}_{k=1}^K} \sum\limits_{k=1}^{K}\frac{\tau}{t} R_k \label{8} \\
  & {\rm s.t.}\  \sum\limits_{k=1}^{K} P_k = P_{\rm sum} , \label{9} \\
  & \qquad 0 \leq P_k \leq P_{\rm sum} , \ \forall \,k , \label{10} \\
  & \qquad \textbf{f}_k^{\rm H}\textbf{f}_k\leq 1, \ \forall \,k , \label{11} \\
  & \qquad \psi_{k,m} = \frac{2\pi a}{2^e-1}, a\in\{0,1,\dots ,2^e-1\},\ \forall \,k, m, \label{12} \\
  & \qquad \big\|\theta_{k, m}\big\| = 1,\ \forall \,k, m. \label{13}
\end{align}
The constraints (\ref{9}) and (\ref{10}) limit the total power consumption to $P_{\rm sum}$, which is the upper bound of power allocated to each cluster. Next (\ref{11}) imposes a magnitude constraint on each beamforming vector, while (\ref{12}) specifies the discrete set from which the phase shift of each element takes value, and (\ref{13}) imposes the unit amplitude constraints on IRS elements.

\section{Throughput Maximum Algorithm}\label{S4}

In the joint optimization problem $\mathcal{P}$, the optimization variables in the objective function are coupled while the constraints in (\ref{13}) are non-convex. Thus there exists no standard method to solve $\mathcal{P}$ directly. We propose a two-stage optimization algorithm by leveraging problem decoupling and alternating optimization. In the first stage, we design the transmit beamforming and optimize IRS phase shifts, while in the second stage we adjust the power allocated to each frame with the optimization results obtained in the first stage, to maximize the total throughput.

\subsection{Stage One: Joint Optimization of Beamforming and IRS Phase Shifts}\label{S4.1}

In this stage, we fix the power allocated to each frame to $\overline{P}=P_{\rm sum}/K$ and denote the overall power allocation by $\overline{\textbf{p}}=\big[\overline{P},\overline{P},\dots,\overline{P}\big]^{\rm T}$. With the power allocation vector $\textbf{p}$ fixed to $\overline{\textbf{p}}$, the optimization $\mathcal{P}$ is simplified to
\begin{align} 
  & \mathcal{P}1: \max_{\{\bm{\Theta}_k\}_{k=1}^K,\{\textbf{f}_k\}_{k=1}^K} \sum\limits_{k=1}^{K} \frac{\tau}{t} R_k \label{OpFp} \\
  & {\rm s.t.}\ \ \textbf{f}_k^{\rm H}\textbf{f}_k\leq 1, \ \forall \,k, \label{CfFp} \\
  & \qquad \varphi_{k,m} \in[0, ~ 2\pi],\ \forall \,k,m, \label{CpFp} \\
  & \qquad \big\|\theta_{k,m}\big\|=1,\ \forall \,k,m. \label{CeFp}
\end{align}
Note that we have relaxed the phase shift of each refractive element to be a continuous variable taking value in $[0, ~ 2\pi]$. To avoid high complexity of the joint optimization over the beamforming and IRS phase shifts, we further decouple $\mathcal{P}1$ into two sub-problems, each involving only one set of variables with the other fixed, and an alternating algorithm is adopted to alternately optimize the transmit beamforming and IRS phase shifts until converges.

\subsubsection{Beamforming Design at BS}\label{S4.1.1}

Given $\textbf{p}=\overline{\textbf{p}}$ and the set of $\{\bm{\Theta}_k\}_{k=1}^K$, the objective function of $\mathcal{P}1$ is monotonically increasing with the aggregate signal-to-noise ratio (SNR) of each cluster. Therefore, the optimization over $\{\textbf{f}_k\}_{k=1}^K$ can be equivalently solved by maximizing the SNR of each cluster, where the optimal beamforming vectors can be obtained by solving the following $K$ sub-problems
\begin{align} 
  & \mathcal{P}1_{\textbf{f}_k}: \max_{\textbf{f}_k} \sum\limits_{i=1}^{N} \frac{\overline{P}\left|\textbf{v}^{\rm{H}}_{k,i} \bm{\Theta}_k \textbf{G}_k \textbf{f}_k\right|^2}{\sigma^2} \label{eq18} \\ 
  & \ \ {\rm s.t.}\ \  \textbf{f}_k^{\rm H}\textbf{f}_k\leq 1 , \label{eq19}
\end{align}
for $1\le k\le K$. Although $\mathcal{P}1_{\textbf{f}_k}$ is non-convex with respect to $\textbf{f}_k$, it can be equivalent to write as
\begin{align} 
  & \max_{\textbf{f}_k}\,\frac{\overline{P}}{\sigma^2}\,\textbf{f}^{\rm{H}}_k \left(\sum\limits_{i=1}^{N}\textbf{G}^{\rm{H}}_k\bm{\Theta}^{\rm{H}}_k\textbf{v}_{k,i} \textbf{v}^{\rm{H}}_{k,i} \bm{\Theta}_k \textbf{G}_k \right) \textbf{f}_k , \label{20} 
\end{align}
which is a hermitian nonnegative definite matrix.
Then the optimal beamforming can be driven by 
\begin{equation}\label{equ:MRT} 
  \textbf{f}_k^{\star} ={\bm{x}_{\max}}, 
\end{equation}
where ${\bm{x}_{\max}}\,(||\bm{x}_{\max}||=1)$ denotes the eigenvector corresponding to the largest eigenvalue of matrix $\left(\sum\limits_{i=1}^{N}\textbf{G}^{\rm{H}}_k\bm{\Theta}^{\rm{H}}_k\textbf{v}_{k,i} \textbf{v}^{\rm{H}}_{k,i} \bm{\Theta}_k \textbf{G}_k \right)$.
\subsubsection{IRS Phase Shift Design}\label{S4.1.2}

Given the power allocation vector $\overline{\textbf{p}}$ and beamforming vectors $\{\textbf{f}_k\}_{k=1}^K$, the optimization of IRS phase shifts can be simplified as
\begin{align} 
  & \mathcal{P}1_{\bm{\Theta}}: \max_{\{\bm{\Theta}_k\}_{k=1}^K} \sum\limits_{k=1}^{K} \frac{\tau}{t} R_k \label{eq21} \\
  & \ \ {\rm s.t.} \ \  \varphi_{k,m} \in[0, ~ 2\pi],\ \forall \,k,m, \label{eq22} \\
  & \qquad \ \ \big\|\theta_{k,m}\big\| = 1 , \ \forall \,k,m . \label{eq23}
\end{align}
Due to the TDMA mode, the above problem can be decomposed into the following $K$ separate sub-problems
\begin{align} 
  &	\mathcal{P}1_{\bm{\Theta}_k}: \max_{\bm{\Theta}_k} \frac{\tau}{t} R_k  \label{opti} \\
  & \ \ {\rm s.t.} \ \  \varphi_{k,m} \in[0, ~ 2\pi],\ \forall \, m , \label{optiB21} \\
  & \qquad \ \ \big\|\theta_{k,m}\big\| = 1 , \ \forall \,m , \label{optiB22}
\end{align}
for $1\le k \le K$.

\emph{2.1)~Continuous Phase Shifts:}
The non-convex unit modulus constraint (\ref{optiB22}) can be relaxed as 
\begin{equation}\label{equ:min_power} 
  \big\|\theta_{k,m}\big\|\leq 1, \ \forall \, m,
\end{equation}
which is convex. In addition, $\big|\textbf{v}^{\rm{H}}_{k,i} \bm{\Theta}_k \textbf{G}_{k} \textbf{f}_k\big|^2$ is a convex quadratic function of $\bm{\Theta}_k$, and by applying SCA, it can be approximated by the first-order Taylor expansion. Specifically, given local point $\bm{\theta}_k^{s}=[e^{\textsf{j}\varphi^s_{k,1}},\dots ,e^{\textsf{j}\varphi^s_{k,M}}]^{\mathrm{T}}$ in the $s$-th iteration and let $\textbf{V}_{k,i}=\mathrm{diag}\left(\textbf{v}^{\rm{H}}_{k,i}\right)$, the following relationship holds

\begin{align}\label{equ:V2VA2} 
  & \sum\limits_{i=1}^{N}\big|\textbf{v}^{\rm{H}}_{k,i} \bm{\Theta}_k \textbf{G}_{k} \textbf{f}_k\big|^2\geq 
-\sum\limits_{i=1}^{N}\big|\bm{\theta}^{s,\rm{H}}_k\textbf{V}_{k,i}\textbf{G}_{k}\textbf{f}_k\big|^2\nonumber \\
&\qquad\qquad+2\Re \bigg\{\sum\limits_{i=1}^{N}\bm{\theta}^{s,{\rm H}}_k(\textbf{V}_{k,i}\textbf{G}_{k}\textbf{f}_k\textbf{f}_k^{\rm H}\textbf{G}_{k}^{\rm H}\textbf{V}_{k,i}^{\rm H})\,\bm{\theta}_k\bigg\},
\end{align}
which is linear and convex. \\
\emph{Proof:} See Appendix B.

As a result, $\mathcal{P}1_{\bm{\Theta}_k}$ can be approximated as a convex problem, which can be solved by an interior-point method \cite{Interior}. Therefore, we can obtain the optimal phase shift vector $\bm{\varphi}^{\star}_k$ for each cluster, with $\bm{\varphi}^{\star}_k=\big[\varphi^{\star}_{k,1}, \varphi^{\star}_{k,2},\dots ,\varphi^{\star}_{k,M}\big]$.
 
\emph{2.2)~Discrete Phase Shifts:}
Since it is challenging to realize continuous amplitude phase-shift values in practice, we quantize the obtained solutions to discrete values within the intervals of length $\Delta =\frac{2\pi}{2^e -1}$. Let $l_m=\big\lfloor\frac{\varphi^{\star}_{k,m}}{\Delta}\big\rfloor$. Then $\varphi^{\star}_{k,m}\in\big[l_m\Delta, ~ (l_m+1)\Delta\big]$. The search space for the $M$ discrete phase shifts comprises $2^M$ possibilities. To select the optimal discrete phase shift set from them, a decision method based on the BB algorithm is proposed. 

\begin{algorithm}[!htp]  
\caption{BB-based Phase Shift Algorithm}
\label{alg:1}  
\KwIn{The optimal continuous phase shift vector $\bm{\varphi}^{\star}_k$, \\
\qquad\quad A feasible solution of discrete phase shift vector \\
\qquad\quad $\bm{\psi}^0_k$ and the corresponding throughput $q_k^L$\; 
}
  \SetKwInput{KwInitialization}{Initialization} 
    \KwInitialization{The optimal discrete phase shift vector $\bm{\psi}^{\star}_k\leftarrow \emptyset$\;}
  \SetKwProg{Function}{function}{}{end}
  \Function{\textsc{Branch(\textit{m})}}{
    \For{$j=0:1$}{
      $\psi_{k,m}=(l_m+j)\Delta$\;
      \eIf{$m=M$}
        {Calculate the corresponding throughput $q_k$\;
        \If{$q_k>q^{L}_k$}
          {$q^{L}_k=q_k$\;
           $\bm{\psi}^{\star}_k$ is the set of all phase shift values at current path\;}
		    }
        {Calculate $q^{U}_k=\max\limits_{\widetilde{\bm{\psi}}_k}\frac{\tau}{t} R_k$\;
        \If{$q^U_k>q^L_k$}
        {Call $\textsc{Branch(\textit{m}+1)}$\;}
		    }
    }
	}
  {$\textsc{Branch(1)}$}\;
  \Return $\bm{\psi}^{\star}_k$, $q^L_k$.
\end{algorithm}

As shown in Algorithm~\ref{alg:1}, we input the optimal continuous phase shift vector $\bm{\varphi}^{\star}_k$, an initial feasible discrete solution $\bm{\psi}^0_k=\big[\psi^0_{k,1},\dots ,\psi^0_{k,M}\big]$ and the corresponding throughput $q^L_k$ as the lower bound of $\mathcal{P}1_{\bm{\Theta}_k}$. Besides, $\bm{\psi}^{\star}_k$ is initialized as an empty set to store the optimal discrete phase shifts. Next, starting from $m=1$, the algorithm calls function $\textsc{Branch(\textit{m})}$ (line~13 of Algorithm~\ref{alg:1}) to traverse feasible discrete phase shifts and find the optimal solution. For current element $m$, $\textsc{Branch(\textit{m})}$ considers its two possible values iteratively. Note that in each loop, the following two cases are considered.
\\
a)~When $m$ arrives at the final element, a complete set of phase shift values is obtained and the corresponding throughput $q_k$ is obtained. Then it is compared with the current lower bound, and if $q_k>q^L_k$, $q^L_k$ is updated to $q^L_k=q_k$ and terminate $\textsc{Branch(\textit{m})}$. \\
b) When $m$ has not reached $M$, we calculate $q^U_k$ (the upper bound of the throughput) by relaxing the phase shifts from $m$ to $M$ to continuous values (denoted as $\widetilde{\bm{\psi}}_k$). If $q^U_k>q^L_k$, call function $\textsc{Branch(\textit{m}+1)}$; otherwise, terminate this path. 

The above procedure iterates until all the possible paths are tried, after which the optimal discrete-valued phase shift vector $\bm{\psi}^{\star}_k$ and the corresponding maximum throughput $q^L_k$ are returned. The worst-case computation complexity of this algorithm is on the order of $\textsf{O}\big(2^M\big)$.

\begin{algorithm}[!htb] 
\caption{Alternating Optimization for Solving $\mathcal{P}1$}
\label{alg:AO} 
\SetKwInOut{Initialization}{Initialization}
\Initialization{$\overline{\textbf{p}}=\big[\overline{P},\overline{P},\dots ,\overline{P}\big]^{\rm T},\textbf{f}^0=\big\{\textbf{f}^0_1,\textbf{f}^0_2,\dots ,\textbf{f}^0_K\big\}$,
$\bm{\psi}^0=\big\{\bm{\psi}^0_1,\bm{\psi}^0_2,\dots ,\bm{\psi}^0_K\big\}$, corresponding throughput $q^0$,
stopping criterion $\delta = 10^{-3}$, iteration count $s=0$;}
\SetKwRepeat{Do}{do}{while}
  \Do{$|q^{s+1}-q^{s}| \geq \delta$}
  {
    \For{$k=1:K$}{
      Update beamforming vector $\textbf{f}^{s+1}_k$ by Eq.\,($\ref{equ:MRT}$) with given $\bm{\psi}^s_k$\;
    }
    \For{$k=1:K$}{
		  Obtain optimal continuous phase shifts using interior-point method \cite{Interior} with given $\textbf{f}^{s+1}_k$\;
      Calculate discrete phase shift vector $\bm{\psi}^{s+1}_k$ by Algorithm~$\ref{alg:1}$\;
		}
    Calculate $q^{s+1}$ with given $\overline{\textbf{p}}$, $\textbf{f}^{s+1}$ and $\bm{\psi}^{s+1}$\;
  }
\Return $\textbf{f}^{\star}=\textbf{f}^s$, $\bm{\psi}^{\star}=\bm{\psi}^s$ and throughput $q^{\star}=q^s$.
\end{algorithm}

\subsubsection{Alternating Optimization}\label{S4.1.3}

Based on the solutions of Subsections~\ref{S4.1.1} and \ref{S4.1.2}, we devise an alternating optimization (AO) method to solve the optimization problem $\mathcal{P}1$, as summarized in Algorithm~\ref{alg:AO}. To begin with, we fix the transmit power to $\overline{\textbf{p}}$, initialize $\textbf{f}^0$ and $\bm{\psi}^0$, besides calculating the corresponding throughput $q^0$. Furthermore, stopping criterion $\delta$ is set and iteration count is initialized to $s=0$. In each loop, $\textbf{f}^{s+1}_k$ of each cluster is calculated by the closed-form solution ($\ref{equ:MRT}$) with fixed $\bm{\psi}^{s}$. Then with fixed $\textbf{f}^{s+1}$, the continuous phase shifts are obtained by optimizing the convex approximation of each $\mathcal{P}1_{\bm{\Theta}_k}$, based on which the discrete $\bm{\psi}^{s+1}$ can be obtained using Algorithm~$\ref{alg:1}$. Lastly, the corresponding throughput is calculated. The iteration procedure is terminated when the throughput difference between two successive iterations is smaller than $\delta$, yielding the optimal solution $\big(\textbf{f}^{\star},\bm{\psi}^{\star},q^{\star}\big)$. The discussions on the convergence and complexity of Algorithm~\ref{alg:AO} is given below.

\emph{3.1)~Convergence:} 
Denote the throughput at the $s$-th iteration as $q^s=q\left(\overline{\textbf{p}}, \textbf{f}^s, \bm{\psi}^{s}\right)$, where $\overline{\textbf{p}}$, $\textbf{f}^s$, $\bm{\psi}^{s}$ are the input of next iteration. As steps 2-3 solve the optimal beamforming vector $\textbf{f}^{s+1}_k$ for each cluster, $q\left(\overline{\textbf{p}}, \textbf{f}^{s}, \bm{\psi}^{s}\right)\leq q\left(\overline{\textbf{p}}, \textbf{f}^{s+1}, \bm{\psi}^{s}\right)$ holds. Then steps 4-6 update the phase shift vector for each cluster into $\bm{\psi}^{s+1}_k$ based on $\textbf{f}^{s+1}$, and the result $q\left(\overline{\textbf{p}}, \textbf{f}^{s+1}, \bm{\psi}^{s+1}\right)$ satisfies the following inequality
\begin{equation}\label{equ:V2VA3} 
  q\big(\overline{\textbf{p}}, \textbf{f}^s, \bm{\psi}^{s}\big) \leq q\big(\overline{\textbf{p}}, \textbf{f}^{s+1}, \bm{\psi}^{s}\big) \leq q\big(\overline{\textbf{p}}, \textbf{f}^{s+1},\bm{\psi}^{s+1}\big) ,
\end{equation} 
which indicates that the objective function is non-decreasing in the consecutive iterations. Since the throughput is upper-bounded by a finite value due to the power constraint, we conclude that Algorithm~\ref{alg:AO} converges.

\emph{3.2)~Complexity:}
In each iteration of Algorithm~\ref{alg:AO}, the computational complexity of beamforming design is negligible. The main computational complexity $\textsf{O}\left(K\left(\mathcal{N}^{3.5}+2^M\right)\right)$ comes from solving the optimal discrete phase shifts (steps 4-6), where $\mathcal{N}$ denotes the number of variables utilized by the interior-point method. As the iteration is conducted for $s$ times, the total complexity of Algorithm~\ref{alg:AO} is $\textsf{O}\left(sK\big(\mathcal{N}^{3.5}+2^M\big)\right)$. 

\subsection{Stage two: Power Allocation and Throughput Maximization}\label{S4.2}

Because the CSI may change significantly over the transmission duration of $t$ or $K$ frames, we execute the power allocation every $l$ frames given beamformers and IRS phase shifts obtained in Stage one. Notice that $l$ is carefully chosen such that the CSI can be predictable over the duration of $l\tau$. Obviously, the power allocation among $l$ clusters can be reduced to the following optimization problem
\begin{align}
  \max_{\textbf{p}_l} & \sum\limits_{k=\rho+1}^{\rho+l} \frac{\tau}{t} \log_2\left( 1 + P_k \sum\limits_{i=1}^{N}\frac{|\textbf{v}^{\rm H}_{k,i} \bm{\Theta}_k \textbf{G}_{k} \textbf{f}_k|^2}{\sigma^2}\right), \label{optiB} \\
  {\rm s.t.} & \sum\limits_{k=\rho+1}^{\rho+l} P_k=l \overline{P},  \label{Ctp} \\
  & 0\leq P_k \leq l \overline{P},\ \rho+1 \le k\le \rho+l , \label{Cip}
\end{align}
where $\textbf{p}_l=\big[P_{\rho+1},\dots ,P_{\rho+l}\big]^{\rm T}$ is the power allocation vector for the current $l$ frames, and $\rho$ is the number of clusters whose power allocations have been executed. Note that we have set the total power consumption for every $l$ frames to $l\overline{P}=\frac{l}{K}P_{\rm sum}$. As (\ref{optiB}) is convex with respect to $P_k$ and all constraints are linear, this optimization problem is convex and can be solved by the Lagrangian multiplier method. Specifically, the Lagrangian function can be expressed as
\begin{align}\label{equ:La}  
  L\left(\textbf{p}_l,\bm{\lambda},\bm{\beta},\mu\right) =& \sum\limits_{k=\rho+1}^{\rho+l} - R_k - \lambda_k P_k- \beta_k\big( l \overline{P} - P_k \big) \nonumber \\
  & -\mu \left(\sum\limits_{k=\rho+1}^{\rho+l} P_k - l \overline{P}\right) , 
\end{align}
where $\bm{\lambda}=\big[\lambda_{\rho+1},\dots ,\lambda_{\rho+l}\big]^{\rm T}$ and $\bm{\beta}=\big[\beta_{\rho+1},\dots ,\beta_{\rho+l}\big]^{\rm T}$ are the Lagrangian multipliers associated with constraints in (\ref{Cip}); $\mu$ is the Lagrangian multiplier associated with  constraint (\ref{Ctp}).

The Karush-Kuhn-Tucker (KKT) condition of (\ref{equ:La}) associated with $P_k$ is given by
\begin{align}\label{equ:KKT} 
  & \frac{\partial{L\left(\textbf{p}_l,\bm{\lambda},\bm{\beta},\mu\right)}}{\partial P_k} = -\frac{\frac{\sum\limits_{i=1}^{N} |\textbf{v}^{\rm H}_{k,i} \bm{\Theta}_k \textbf{G}_{k} \textbf{f}_k|^2}{\sigma^2}}{\ln 2 \left(1 + P_k \frac{\sum\limits_{i=1}^{N} |\textbf{v}^{\rm H}_{k,i} \bm{\Theta}_k \textbf{G}_{k} \textbf{f}_k|^2}{\sigma^2}\right)} \nonumber \\
  &\hspace*{10mm} - \lambda_k + \beta_k - \mu = 0 , \, \rho+1 \le k\le \rho+l .
\end{align}
An iterative gradient descent procedure can be utilized to obtain the optimal solutions $P_k^{\star}$, $\lambda_k^{\star}$, $\beta_k^{\star}$ and $\mu^{\star}$ based on the KKT conditions of (\ref{equ:La}) associated with $P_k$, $\lambda_k$, $\beta_k$ and $\mu$. Let $\lambda_k^r$, $\beta_k^r$ and $\mu_k^r$ be the Lagrangian multiplier solutions after the $r$-th iteration. From (\ref{equ:KKT}), the power allocation solution $P_k^r$, $\rho+1 \le k\le \rho+l$, after the $r$-th iteration, can be obtained in the closed-form as
\begin{align}\label{equ:P^*_i(k)} 
 P_k^r =& \frac{1}{\ln2 \big(\beta_k^r - \lambda_k^r - \mu^r \big)} - \frac{\sigma^2}{\sum\limits_{i=1}^{N}\big|\textbf{v}^{\rm H}_{k,i}\bm{\Theta}_k\textbf{G}_{k}\textbf{f}_k\big|^2} .
\end{align}
Then the Lagrangian multipliers are updated according to the associated KKT conditions as
\begin{align} 
  \lambda^{r+1}_k =& \lambda^{r}_k - c^{r}_{k} P_k^r , \, \rho+1 \le k\le \rho+l \label{eq36}  \\
  \beta^{r+1}_k =& \beta^{r}_k - d^{r}_{k} \big( l \overline{P} - P_k^r\big) , \, \rho+1 \le k\le \rho+l \label{eq37} \\
  \mu^{r+1} =& \mu^{r} - e^{r} \left(\sum\limits_{k=\rho+1}^{\rho+l} P_k^r - l \overline{P}\right) , \label{eq38}
\end{align}
where $c_k^r$, $d_k^r$ and $e^r$ denote the corresponding step sizes.

\begin{algorithm}[!htb] 
\caption{Power Allocation for Throughput Maximization}
\label{alg:3} 
\SetKwInOut{Input}{Input}
\Input{Optimal beamformers and IRS phase shifts of the users in every $l$ clusters $\textbf{f}^{\star}$ and $\bm{\psi}^{\star}$;}
\SetKwInOut{Initialization}{Initialization}
\Initialization{Maximum iteration times $r_{\max}=100$, $\bm{\lambda}^0$, $\bm{\beta}^0$, $\bm{c}^0$, $\bm{d}^0$, $\mu^0=50$, $e^0=10$, stopping criterion $\varepsilon = 10^{-2}$, iteration index $r=0$;}
  \While{$r \leq r_{\max}$}{
    Calculate $\textbf{p}_l^{r}$ with (\ref{equ:P^*_i(k)})\; 
    Update $\bm{\lambda}^r$, $\bm{\beta}^r$, $\mu^r$ with (\ref{eq36})-(\ref{eq38})\;
    \If{any of $c^r_k, d^r_k, e^r>1$}{
      Decrease its value by half in the next iteration\;
		}
    \If{$\big\|\bm{\lambda}^{r+1} - \bm{\lambda}^{r}\big\| < \varepsilon$ and $\big\|\bm{\beta}^{r+1} - \bm{\beta}^{r}\big\| < \varepsilon$ and $\big|\mu^{r+1}-\mu^{r}\big| < \varepsilon$}{
      break\;
		}
		$r=r+1$\;
  }
	Calculate the optimal throughput $q\big(\textbf{p}_l^{\star}=\textbf{p}_l^r,\textbf{f}^{\star},\bm{\psi}^{\star}\big)$\;
\Return $\textbf{p}_l^{\star}$, $q\big(\textbf{p}_l^{\star},\textbf{f}^{\star},\bm{\psi}^{\star}\big)$.
\end{algorithm} 

Algorithm~\ref{alg:3} summarizes the proposed power allocation procedure for throughput maximization based on the Lagrangian multiplier method. For notation convenience, we still denote the optimal beamformers and IRS phase shifts of the users in every $l$ clusters obtained in Stage One by $\textbf{f}^{\star}$ and $\bm{\psi}^{\star}$, which together with the maximum iteration times $r_{\max}$ and stopping criterion $\varepsilon$ are inputted to the algorithm. Also, the initial values of Lagrangian multipliers and the corresponding gradient descent step sizes are given. In each iteration, each $P_k$ is calculated by (\ref{equ:P^*_i(k)}) in line~2, and the Lagrangian multipliers are updated according to (\ref{eq36})-(\ref{eq38}) in line~3. Then the step sizes are adjusted appropriately in lines 4-5. After convergence, the optimal power allocation vector $\textbf{p}_l^{\star}$ and the corresponding throughput $q\big(\textbf{p}_l^{\star},\textbf{f}^{\star},\bm{\psi}^{\star}\big)$ are returned.

Observe that the step sizes in Algorithm~\ref{alg:3} remain positive and non-increasing over the iteration procedure. More importantly, due to the convexity of  (\ref{optiB})-(\ref{Cip}), the KKT point is also the global optimal solution in the $l$ frames. Executing this algorithm for $\frac{K}{l}$ times, we obtain the total power allocation for the entire transmission period.

\section{Performance Evaluation}\label{S5}

We provide numerical simulation results to verify the effectiveness of our proposed IRS-assisted throughput maximization approach for mmWave HST communication systems. 

\begin{table}[!b] 
\vspace*{-4mm}
\caption{Default System Parameters}
\label{table:Initial_Parameters} 
\vspace*{-4mm}
\begin{center}
\begin{tabular}{lcl}  
\toprule   
  Parameter                    & Symbol         & Value \\ \midrule 
 Carrier frequency            & $f$            & $28$ GHz \\
   Frame duration               & $\tau$         & $36$ ms  \\
 
  Coverage radius of the cell  & $R$            & $350$ m  \\
  Distance between BS and rail & $D_0$          &$20$ m \\
  Average transmit power       & $\overline{P}$ & $20$ dBm \\
  Number of BS antennas        & $L$            & $16$ \\
  IRS elements                 & $M$            & $64$ \\
  System bandwidth             & $W$            & $2000$ MHz \\
  Background noise             & $N_0$          & $-80$ dBm/Hz \\
  Rician K-factor              & $K_f$          & $3$ dB \\
  PL factor of BS-IRS links    & $\beta_1$      & $2$ dB \\
  PL factor of IRS-user links  & $\beta_2$      & $3$ dB \\
  Speed of HST                 & $v$            & $300$ km/h \\
  Length of power allocation   & $l$            & $10$ \\
	Phase shift quatization bits & $e$            & $2$ \\
	
	Number of users served  & $I$            &  $92$  \\
  \bottomrule  
\end{tabular}
\end{center}
\vspace*{-1mm}
\end{table}

\subsection{Simulation Setup}\label{S5.1}
Based on the system model of Section III, we consider a $28$\,GHz HST communication system with the signal attenuation $h_0\! =\! -61.3849$\,dB at a reference distance of $1$\,m \cite{VC, VC2, WL}. The train consists of ten carriages, each of $24$\,m length, $5$\,m width and $2.5$\,m height \cite{trainsize}. An IRS is assumed to be deployed on the window of the first carriage, which is a uniform square array of $64$ elements with half-wavelength spacing to assist T2G communications of randomly distributed users. Set the number of BS transmit antennas as $16$ with half-wavelength antenna spacing. The frame duration is $36$\,ms and the BS coverage radius is $350$\,m \cite{LPP, CC}, where the time resource allocated to each carriage is $23$ frames due to the TDMA mode. Referring to \cite{901, 902}, there are around $90$ seats per carriage, so it is reasonable to set $N=4$ in the simulation. 
For ease of description, the locations of communication nodes are represented by three-dimensional (3D) coordinates, where the BS is fixed at $(20, 0, 2)$;   the IRS is initially located at $(0, 0, 1)$ and varies on the YZ plane. In this case, we update the positions of the IRS and users frame by frame to accurately reflect the mobility process. What's more, as reference \cite{length} have illustrated that the CSI prediction duration in the mmWave HST scenario can achieve $1.25$\,s, it is practical to define $l=10$ and $l\tau=360$ ms for local power allocation, which balances the prediction complexity and power allocation accuracy. 
Unless otherwise stated, the simulation parameters used are listed in Table~\ref{table:Initial_Parameters}.

Four existing schemes are utilized for performance comparison with our proposed approach, which include
\begin{itemize}
\item \textbf{Neighbor-based Cross-entropy (NCE)} \cite{VC2}: This low-complexity algorithm selects phase shifts of the IRS in the candidate set, which is generated based on the probability distribution function and extended through neighbor extraction. NCE is one of the state-of-the-art schemes to jointly optimize the precoding at the BS and phase shifts at the IRS, also applicable to irregular RIS systems. 
\item \textbf{Successive Refinement (SR)} \cite{Omni, WTCOM}: This scheme considers the suboptimal zero-forcing (ZF) beamforming at the BS and alternately optimizes the phase shift of each programmable element by fixing the others. To the best of our knowledge, the SR algorithm outperforms solutions based on the semi-infinite relaxation (SDR) technique and has received attention in recent studies, therefore considered as a practical benchmark.
\item \textbf{Random Phase Shift (RPS)}: This scheme randomly selects a feasible phase shift for each IRS element, being an important benchmark in many representative RIS-related works \cite{HM2021, RIS2022TVT, RIS2023L, RISTVT3}.
\item \textbf{Without  IRS}: This case does not consider the assistance of IRS, and on-board users can only receive signals going through the carriage.  Comparison with this scheme allows us to validate the theoretical analysis and demonstrate the effectiveness of IRS with discrete phase shifts for improving the performance of HST networks \cite{WTCOM}. 
\end{itemize}

All the simulation results are averaged over 200 independent experiments.

\begin{figure}[!b]
\vspace*{-3mm}
\begin{center}
  \includegraphics[width=\columnwidth]{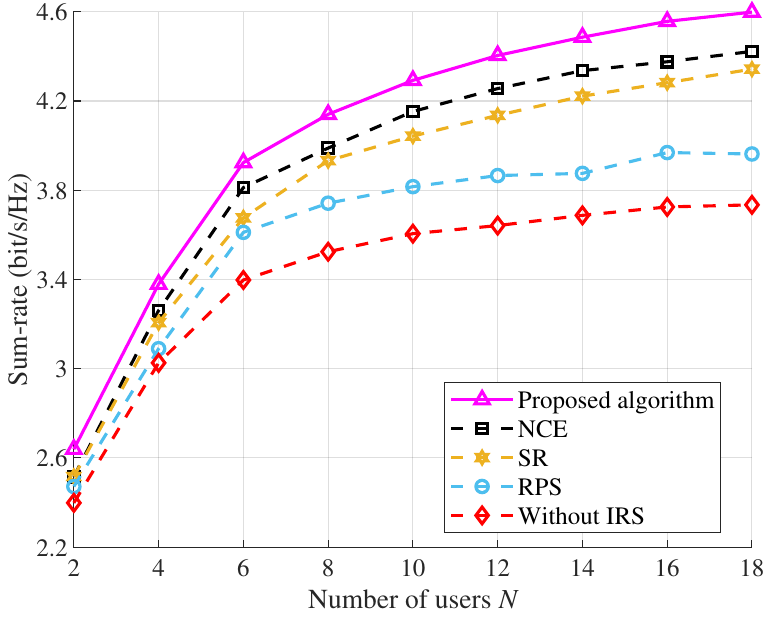}
\end{center}
\vspace*{-5mm}
\caption{Sum rate versus the number of users per frame.}
\label{fig:user} 
\vspace*{-1mm}
\end{figure}

\subsection{The effect of IRS adoption}\label{S5.2}

In this subsection, we evaluate the sum-rate of users in one frame to demonstrate the effect of IRS deployment and compare the performance of five schemes by varying the number of users, the number of IRS elements, the K-factor and the quantification bits $e$, respectively. In experiments, users are distributed randomly in the carriage. Note that the sum-rate is the throughput in one frame.

In Fig.~\ref{fig:user}, we plot the sum-rate achieved by these five schemes as a function of the number of users per cluster $N$. From the results, we observe that the sum rate increases logarithmically with $N$, and our scheme outperforms the other four schemes. Specifically, as more users to be served, the gap between the proposed algorithm and the NCE scheme is increasing, which shows that the neighbor extraction method can hardly find the optimal phase shifts. Then, the proposed scheme attains a significantly higher sum rate compared to the classic SR. This is because 1) alternately optimizing the phase shift of each element is inferior to jointly optimizing all phase shifts and 2) ZF beamforming is not optimal. Besides, compared to the case without IRS, the RPS scheme only achieves a slight performance improvement, which demonstrates the necessity of phase shift optimization.
Finally, at $N=18$, the performance gains of our scheme over the NCE, SR, RPS and Without IRS are around $4.5$\%, $5.7$\%, $16.5$\% and $23.7$\%, respectively.

\begin{figure}[!t]
\begin{center}
  \includegraphics[width=\columnwidth]{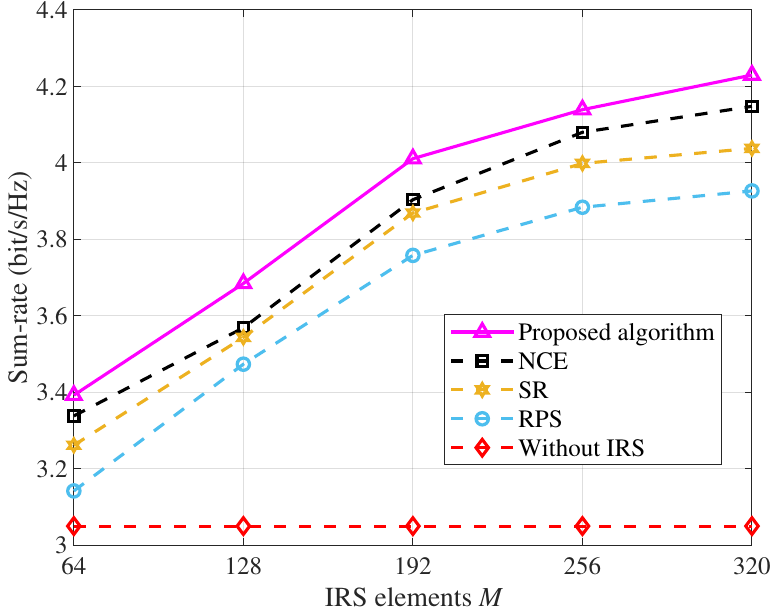}
\end{center}
\vspace*{-5mm}
\caption{Sum rate versus the number of IRS elements.}
\label{fig:IRS Size} 
\vspace*{-3mm}
\end{figure}
Fig.~\ref{fig:IRS Size} illustrates the influence of IRS size on the throughput performance of these five schemes, by varying the number of IRS elements $M$ from $64$ to $320$. As expected, the four schemes with IRS significantly outperform the scheme without IRS, and their sum-rate performance increase with $M$, since more IRS elements result in sharper beams which refract incident signals towards desired devices and mitigate interferences among users more effectively. Again our scheme achieves the best performance. It can be seen from Fig.~\ref{fig:IRS Size} that the performance gap between our proposed scheme and the second-best NCE is 0.15\,bit/s/Hz to 0.1\,bit/s/Hz for $128 \le M\le 320$. But the NCE algorithm offers the advantage of low complexity.

\begin{figure}[!h]
\begin{center}
  \includegraphics[width=\columnwidth]{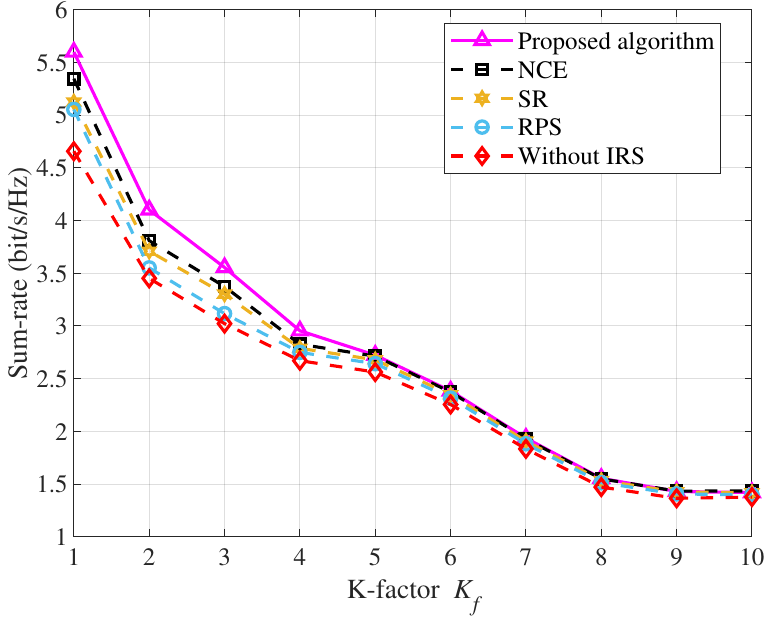}
\end{center}
\vspace*{-5mm}
\caption{Sum rate versus K-factor.}
\label{fig:Kfactor} 
\end{figure}

\begin{figure}[!t]
\begin{center}
  \includegraphics[width=\columnwidth]{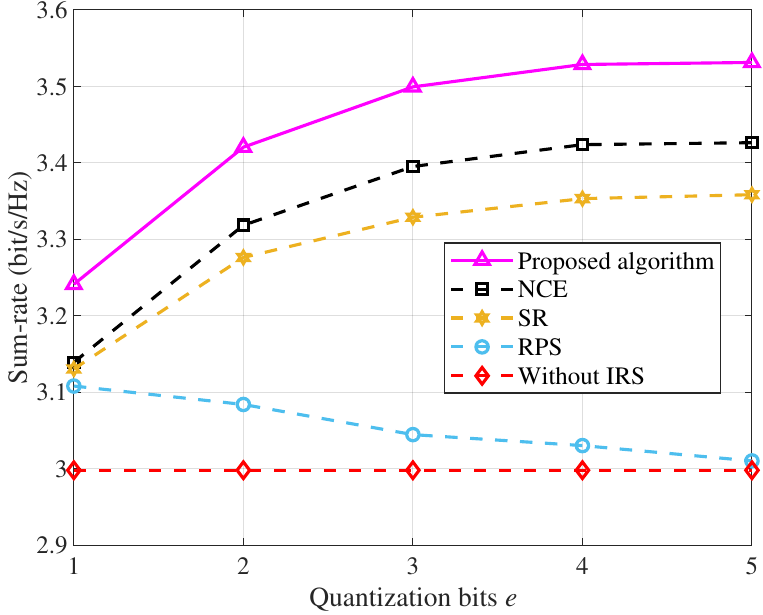}
\end{center}
\vspace*{-5mm}
\caption{Sum rate versus quantification bits.}
\label{fig:Quantification bits} 
\vspace*{-3mm}
\end{figure}

Fig.~\ref{fig:Kfactor} shows the impact of Rician K-factor on the achievable sum-rates of the five schemes. Observe that the sum rate is a decreasing function with respect to $K_f$, and the performance gap among the five schemes becomes negligible when $K_f \geq 5$. This is because as $K_f$ increases, the LoS path becomes stronger while the adjustable multipaths decrease, thus limiting the capacity improvement obtained through IRS deployment. Considering the characteristics of railway communication, $K_f=3$ or $4$ is typical in practice \cite{LGK}.

Fig.~\ref{fig:Quantification bits} compares the sum-rates of the five schemes at different quantization bits $e$. As expected, the proposed algorithm attains the best performance and the curve without IRS stays flat. Moreover, the achievable rates of the proposed, NCE and SR schemes first increase with $e$ and then reach their respective saturation values of 3.53\,bit/s/Hz, 3.42\,bit/s/Hz and 3.36\,bit/s/Hz when the number of quantization bits $e$ exceeds $4$. By contrast, the performance of the RPS decreases as $e$ increases, and becomes close to that of the scheme without IRS at $e = 5$. The reason is that RPS adjusts the phase shift of each element in a completely random manner, and the adjusted signal is not necessarily stronger than the incident signal. Evidently, IRS deployment with suitable optimization and quantization bits enables higher network throughput.
\begin{figure}[!htb]
\vspace*{-1mm}
\begin{center}
  \includegraphics[width=\columnwidth]{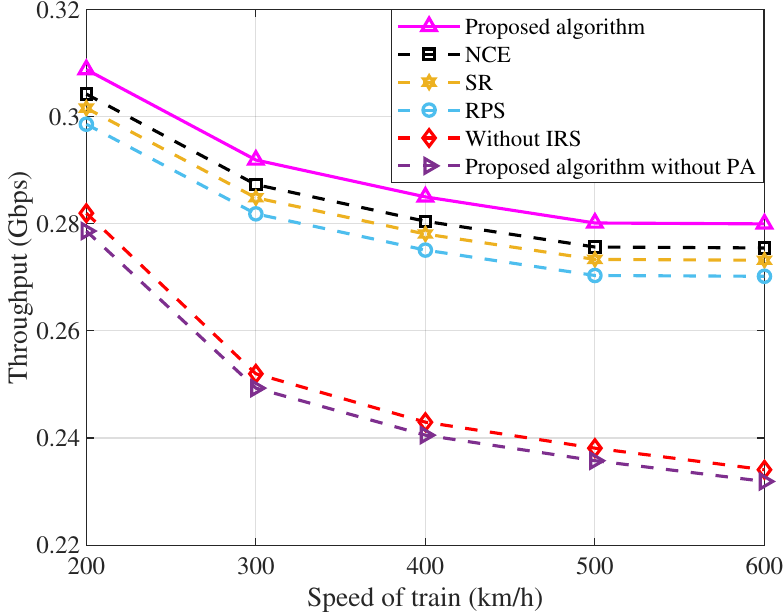}
\end{center}
\vspace*{-5mm}
\caption{Achievable throughput versus mobility speed.}
\label{fig:7}
\vspace*{-1mm}
\end{figure}
\subsection{The overall throughput}

Next we evaluate the achievable overall throughput of the proposed algorithm, the NCE, SR and RPS schemes as well as the scheme without IRS. Fig.~\ref{fig:7} depicts the average throughput of these five schemes as a function of the HST speed $v$ in time duration $t$, given the default system parameters of $N=4$, $M=64$, $e=2$ and $K_f=3$. As expected, the throughput of all five schemes decreases as $v$ increases, because the communication distance grows with the speed for the same time duration. Notice the proposed scheme attains the best performance.

As explained in Section~\ref{S4}, our proposed scheme consists of two stages, with stage one performing joint optimization of beamforming and IRS phase shifts, stage two carrying out power allocation for throughput maximization based on the results of stage one. The proposed power allocation, namely, Algorithm~\ref{alg:3}, is crucial for maximizing the throughput. To demonstrate the effectiveness of Algorithm~\ref{alg:3}, in Fig.~\ref{fig:7}, we also plot the average throughput achieved by performing only the stage-one optimization with equal power allocation, denoted as the proposed algorithm without power allocation. It can be seen that the proposed power allocation algorithm significantly improves the system throughput, and it outperforms the scheme without power allocation by $10.4$\% at $v=200$\,km/h and by $20.7$\% at $v=600$\,km/h.

\section{Conclusions}\label{S6}

In this paper, we have considered the downlink communication in an IRS-assisted mmWave HST system, where user clustering and hybrid TDMA-NOMA technique are leveraged to mitigate transmission interference. To improve the communication quality, we formulate the throughput maximization problem and propose a two-stage solution. In stage one, beamforming and phase shifts are alternately optimized frame by frame. Specifically, given fixed phase shifts, the optimal beamforming is enabled, after which phase shifts are optimized with SCA and the BB algorithm. Then in stage two, based on the results of stage one, the local power allocation is performed to maximize the overall throughput. Numerical simulation results have verified the effectiveness of IRS adoption and confirmed the superior performance of the proposed algorithm over existing solutions. Furthermore, the results have demonstrated that the proposed power allocation algorithm in stage two significantly improves the system throughput. 
In future work, we will investigate the throughput optimization scheme under non-uniform user grouping between frames, besides the effect of train size on system performance.

\begin{appendices}
\section{}
Substituting (6) into (7), we obtain
\begin{small}
\begin{align}\label{equ:R}  
R_k =& \sum\limits_{i=1}^{N} \log_2\left(1 + \gamma_{k,i} \right) \nonumber \\
=&\log_2\left(1 + \frac{P_k\left|\textbf{v}^{\rm H}_{k,1} \bm{\Theta}_k \textbf{G}_k \textbf{f}_k\right|^2}{\sum\limits_{j=2}^{N} P_k\left|\textbf{v}^{\rm H}_{k,j} \bm{\Theta}_k \textbf{G}_{k} \textbf{f}_k\right|^2 + \sigma^2} \right)\nonumber \\
&\qquad+\log_2\left(1 + \frac{P_k\left|\textbf{v}^{\rm H}_{k,2} \bm{\Theta}_k \textbf{G}_k \textbf{f}_k\right|^2}{\sum\limits_{j=3}^{N} P_k\left|\textbf{v}^{\rm H}_{k,j} \bm{\Theta}_k \textbf{G}_{k} \textbf{f}_k\right|^2 + \sigma^2} \right)+\cdots \nonumber \\ 
&\qquad\qquad\qquad\qquad\quad+\log_2\left(1 + \frac{P_k\left|\textbf{v}^{\rm H}_{k,N} \bm{\Theta}_k \textbf{G}_k \textbf{f}_k\right|^2}{\sigma^2} \right) \nonumber \\
=& \log_2\left(\frac{\sum\limits_{i=1}^{N}P_k\left|\textbf{v}^{\rm H}_{k,i} \bm{\Theta}_k \textbf{G}_k \textbf{f}_k\right|^2+ \sigma^2}{\sum\limits_{j=2}^{N} P_k\left|\textbf{v}^{\rm H}_{k,j} \bm{\Theta}_k \textbf{G}_{k} \textbf{f}_k\right|^2 + \sigma^2}\right . \nonumber \\
\times& \left .\frac{\sum\limits_{i=2}^{N}P_k\left|\textbf{v}^{\rm H}_{k,i} \bm{\Theta}_k \textbf{G}_k \textbf{f}_k\right|^2+ \sigma^2}{\sum\limits_{j=3}^{N} P_k\left|\textbf{v}^{\rm H}_{k,j} \bm{\Theta}_k \textbf{G}_{k} \textbf{f}_k\right|^2 + \sigma^2} \times \cdots \times \frac{P_k\left|\textbf{v}^{\rm H}_{k,N} \bm{\Theta}_k \textbf{G}_k \textbf{f}_k\right|^2+ \sigma^2}{\sigma^2} \right)\nonumber \\
=& \log_2\left(1 + \sum\limits_{i=1}^{N} \frac{P_k\left|\textbf{v}^{\rm H}_{k,i} \bm{\Theta}_k \textbf{G}_k \textbf{f}_k\right|^2}{\sigma^2}\right).
\end{align}
\end{small}

\section{}

\begin{small}
\begin{align}\label{equ:V2VA2} 
  & \sum\limits_{i=1}^{N}\big|\textbf{v}^{\rm{H}}_{k,i} \bm{\Theta}_k \textbf{G}_{k} \textbf{f}_k\big|^2\geq \sum\limits_{i=1}^{N}\big|\bm{\theta}^{s,\rm{H}}_k\textbf{V}_{k,i}\textbf{G}_{k}\textbf{f}_k\big|^2\nonumber \\ 
&\qquad\qquad+ \sum\limits_{i=1}^{N}\left.\frac{\partial (\bm{\theta}_k^{\rm H}\textbf{V}_{k,i}\textbf{G}_{k}\textbf{f}_k\textbf{f}_k^{\rm H}\textbf{G}_{k}^{\rm H}\textbf{V}_{k,i}^{\rm H}\bm{\theta}_k)}{\partial \bm{\theta}_k}\right|_{\bm{\theta}_k=\bm{\theta}^{s}_k}(\bm{\theta}_k-\bm{\theta}^{s}_k)\nonumber \\ 
&=\sum\limits_{i=1}^{N}\big|\bm{\theta}^{s,\rm{H}}_k\textbf{V}_{k,i}\textbf{G}_{k}\textbf{f}_k\big|^2\nonumber \\ 
&\qquad\qquad\qquad\,\,\,+2\Re \left\{\sum\limits_{i=1}^{N}\bm{\theta}^{s,\rm{H}}_k(\textbf{V}_{k,i}\textbf{G}_{k}\textbf{f}_k\textbf{f}_k^{\rm H}\textbf{G}_{k}^{\rm H}\textbf{V}_{k,i}^{\rm H})\cdot(\bm{\theta}_k-\bm{\theta}^{s}_k)\right\} \nonumber \\ 
&=-\sum\limits_{i=1}^{N}\big|\bm{\theta}^{s,\rm{H}}_k\textbf{V}_{k,i}\textbf{G}_{k}\textbf{f}_k\big|^2+2\Re \bigg\{\sum\limits_{i=1}^{N}\bm{\theta}^{s,{\rm H}}_k(\textbf{V}_{k,i}\textbf{G}_{k}\textbf{f}_k\textbf{f}_k^{\rm H}\textbf{G}_{k}^{\rm H}\textbf{V}_{k,i}^{\rm H})\,\bm{\theta}_k\bigg\},
\end{align}

where
\begin{align}
\textbf{V}_{k,i}=\mathrm{diag}\left(\textbf{v}^{\rm{H}}_{k,i}\right),
\end{align}
\vspace{-5mm}
\begin{align}
 \bm{\theta}_k^{s}=[e^{\textsf{j}\varphi^s_{k,1}},\dots ,e^{\textsf{j}\varphi^s_{k,M}}]^{\mathrm{T}}.
\end{align}
\end{small}

\end{appendices}
\bibliographystyle{IEEEtran}

\vskip -2\baselineskip plus -1fil
\begin{IEEEbiography}[{\includegraphics[width=1in,height=1.25in,clip,keepaspectratio]{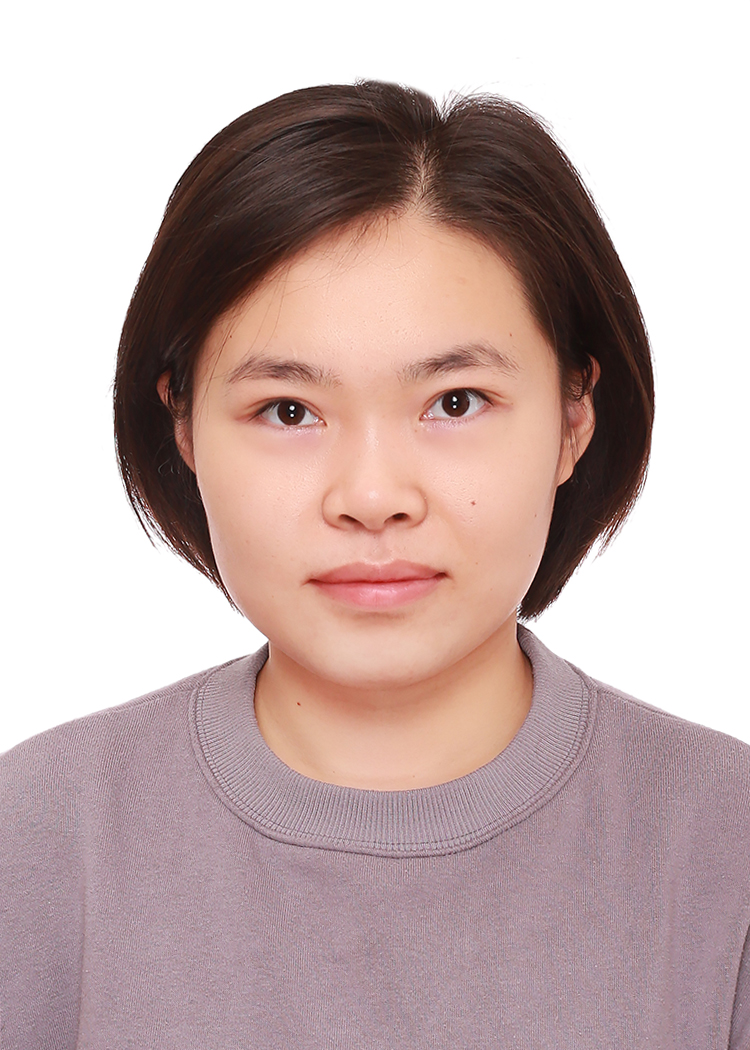}}]{Jing~Li} received the M.E. degree from Beijing Jiaotong University, China, in 2020, where she is currently pursuing her Ph.D. degree. Her research interests include millimeter wave communications, medium access control and reconfigurable intelligent surfaces. 
\end{IEEEbiography}

\vskip -2\baselineskip plus -1fil
\begin{IEEEbiography}[{\includegraphics[width=1in,height=1.25in,clip,keepaspectratio]{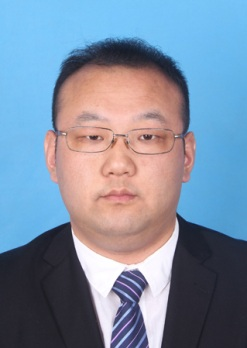}}]{Yong Niu} (IEEE Senior Member) received the B.E. degree in Electrical Engineering from Beijing Jiaotong University, China in 2011, and the Ph.D. degree in Electronic Engineering from Tsinghua University, Beijing, China, in 2016. His research interests are in the areas of networking and communications, including millimeter wave communications, device-to-device communication, medium access control, and software-defined networks. He is currently an Associate Professor with the State Key Laboratory of Rail Traffic Control and Safety, Beijing Jiaotong University. During November 2014 to April 2015, he visited University of Florida, FL, USA as a Visiting Scholar. He received the Ph.D. National Scholarship of China in 2015, Outstanding Ph. D Graduates and Outstanding Doctoral thesis of Tsinghua University in 2016, and Outstanding Ph. D Graduates of Beijing in 2016. He has served as Technical Program Committee (TPC) member for CHINACOM 2015 and IWCMC 2017, and also session chair for IWCMC 2017 and CHINACOM 2017.
\end{IEEEbiography}

\vskip -2\baselineskip plus -1fil
\begin{IEEEbiography}[{\includegraphics[width=1in,height=1.25in,clip,keepaspectratio]{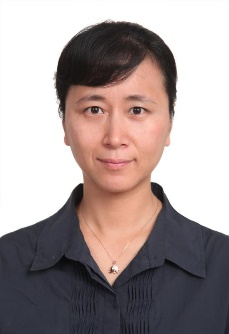}}]{Hao Wu} (IEEE Member) received the Ph.D. degree in information and communication engineering from Harbin Institute of Technology in 2000. She is currently a full professor with the State Key Lab of Rail Traffic Control and Safety at Beijing Jiaotong University (BJTU), China. She has published more than 100 papers in international journals and conferences. Her research interests include Intelligent Transportation Systems (ITS), security and QoS issues in wireless networks (VANETs, MANETs and WSNs), wireless communications, and Internet of Things (IoT). She is a member of IEEE and a reviewer of its major conferences and journals in wireless networks and security.
\end{IEEEbiography}

\vskip -2\baselineskip plus -1fil
\begin{IEEEbiography}[{\includegraphics[width=1in,height=1.25in,clip,keepaspectratio]{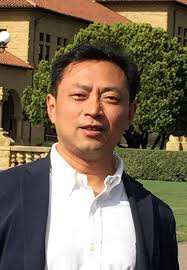}}]{Bo Ai} (IEEE Fellow) received his M.S. and Ph.D. degrees from Xidian University, China, in 2002 and 2004, respectively. He was a visiting professor at EE Department, Stanford University in 2015. He is a full professor and Ph.D. degree candidate advisor with the State Key Laboratory of Advanced Rail Autonomous Operation at Beijing Jiaotong University, China. He has authored/co-authored 8 books and published over 300 academic research papers. He holds 26 invention patents. He is an Institution of Engineering and Technology fellow. He is an associate editor of IEEE Transactions on Consumer Electronics and an editorial committee member of Wireless Personal Communications.
\end{IEEEbiography}

\vskip -2\baselineskip plus -1fil
\begin{IEEEbiography}[{\includegraphics[width=1in,height=1.25in,clip,keepaspectratio]{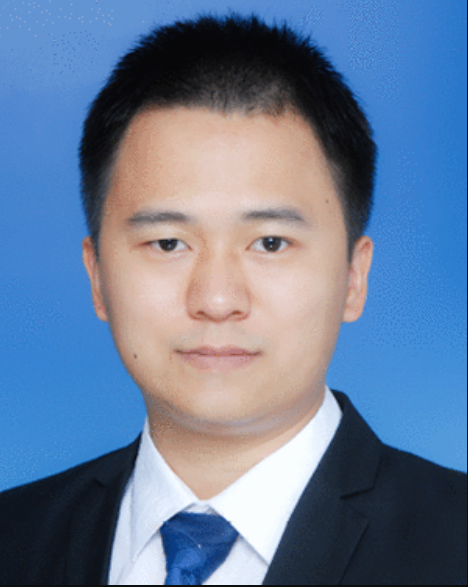}}]{Ruisi He} (IEEE Senior Member)  received the B.E. and Ph.D. degrees from Beijing Jiaotong University (BJTU), Beijing, China, in 2009 and 2015, respectively.
Since 2015, he has been with the State Key Laboratory of Advanced Rail Autonomous Operation, BJTU, where he has been a Full Professor since 2018. He is a Visiting Scholar with the Georgia Institute of Technology, Atlanta, GA, USA, University of Southern California, Los Angeles, CA, USA, and Université Catholique de Louvain, Belgium. He has authored or coauthored five books, three book chapters, more than 200 journal and conference papers, and also several patents. His research interests include wireless propagation channels, railway and vehicular communications, 5G and 6G communications.
He is the Editor of the IEEE Transactions on Wireless Communications, the IEEE Antennas and Propagation Magazine, the IEEE Communications Letters, the IEEE Open Journal of Vehicular Technology, and a Lead Guest Editor of the IEEE Journal on Selected Area in Communications and the IEEE Transactions on Antennas and Propagation. He serves as the Early Career Representative (ECR) of Commission C, International Union of Radio Science (URSI). He was the recipient of the URSI Issac Koga Gold Medal in 2020, the IEEE ComSoc Asia-Pacific Outstanding Young Researcher Award in 2019, the URSI Young Scientist Award in 2015, and five best paper awards in conferences. He is a Member of the COST.
\end{IEEEbiography}

\vskip -2\baselineskip plus -1fil
\begin{IEEEbiography}[{\includegraphics[width=1in,height=1.25in,clip,keepaspectratio]{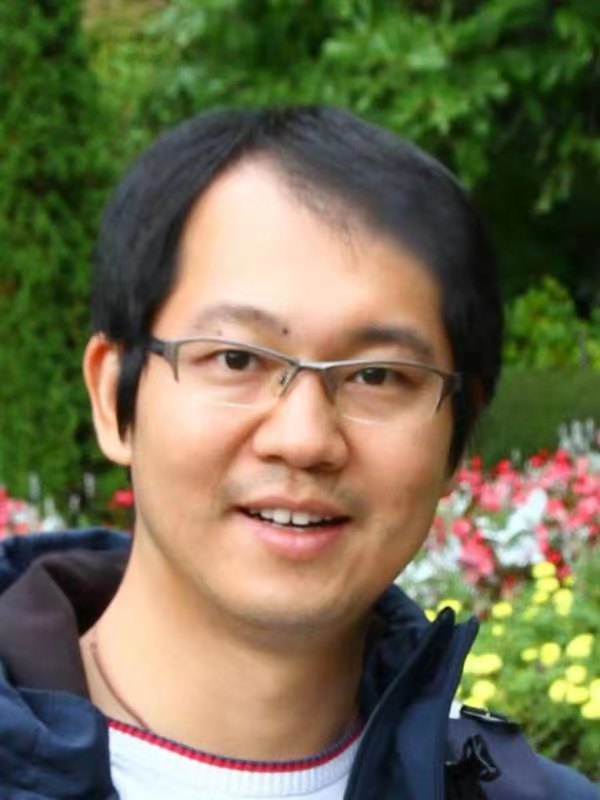}}]{Ning Wang} (IEEE Member) received the B.E. degree in communication engineering from Tianjin University, Tianjin, China, in 2004, the M.A.Sc. degree in electrical engineering from The University of British Columbia, Vancouver, BC, Canada, in 2010, and the Ph.D. degree in electrical engineering from the University of Victoria, Victoria, BC, Canada, in 2013. From 2004 to 2008, he was with the China Information Technology Design and Consulting Institute, as a Mobile Communication System Engineer, specializing in planning and design of commercial mobile communication networks, network traffic analysis, and radio network optimization. From 2013 to 2015, he was a Postdoctoral Research Fellow with the Department of Electrical and Computer Engineering, The University of British Columbia. Since 2015, he has been with the School of Information Engineering, Zhengzhou University, Zhengzhou, China, where he is currently an Associate Professor. He also holds adjunct appointments with the Department of Electrical and Computer Engineering, McMaster University, Hamilton, ON, Canada, and the Department of Electrical and Computer Engineering, University of Victoria, Victoria, BC, Canada. His research interests include resource allocation and security designs of future cellular networks, channel modeling for wireless communications, statistical signal processing, and cooperative wireless communications. He has served on the technical program committees of international conferences, including the IEEE GLOBECOM, IEEE ICC, IEEE WCNC, and CyberC. He was on the Finalist of the Governor Generals Gold Medal for Outstanding Graduating Doctoral Student with the University of Victoria in 2013.
\end{IEEEbiography}

\vskip -2\baselineskip plus -1fil
\begin{IEEEbiography}[{\includegraphics[width=1in,height=1.25in,clip,keepaspectratio]{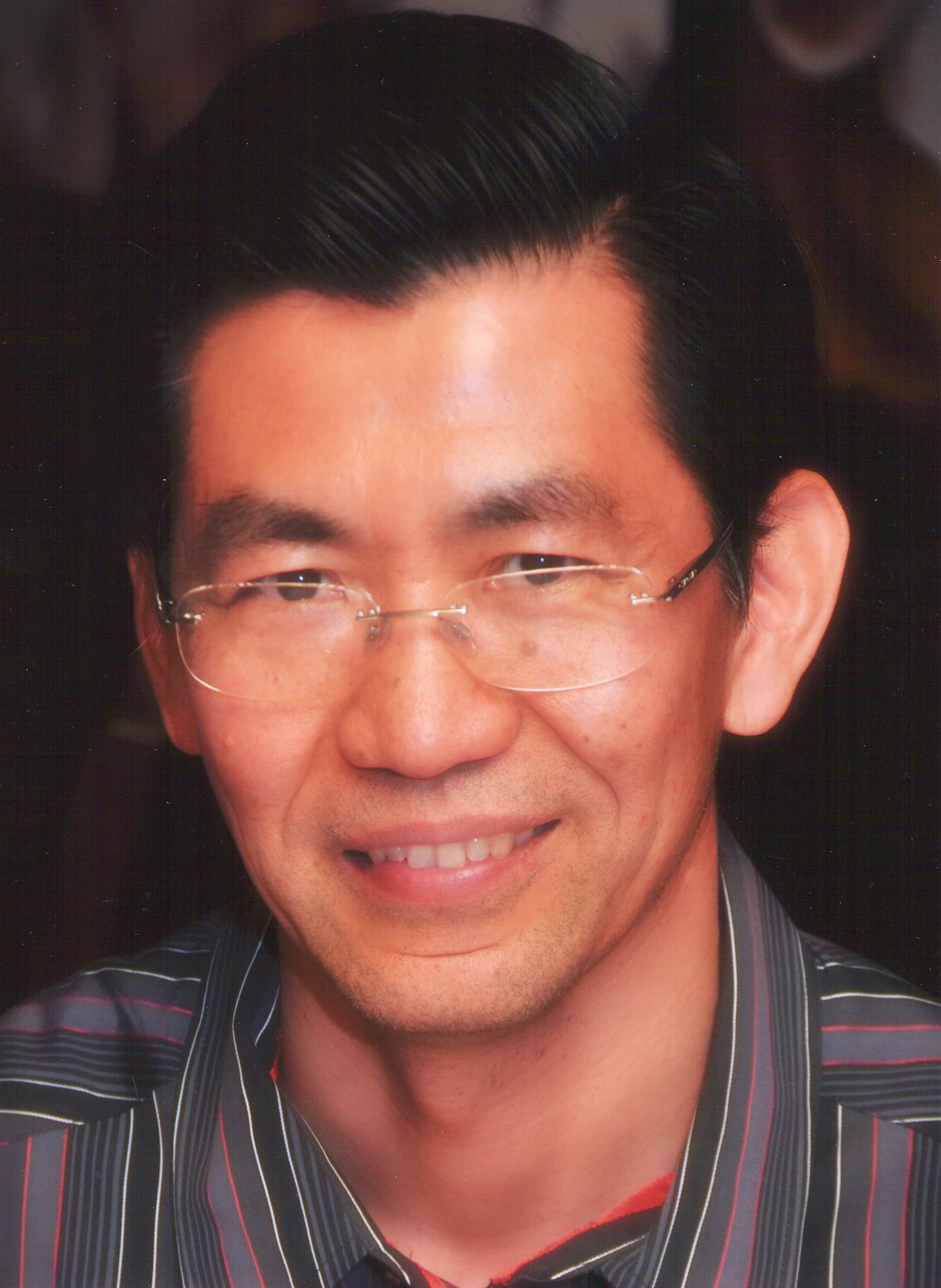}}]{Sheng Chen}
(IEEE Life Fellow) received his BEng degree from the East China Petroleum Institute, Dongying, China, in 1982, and his PhD degree from the City University, London, in 1986, both in control engineering. In 2005, he was awarded the higher doctoral degree, Doctor of Sciences (DSc), from the University of Southampton, Southampton, UK. From 1986 to 1999, He held research and academic appointments at the Universities of Sheffield, Edinburgh and Portsmouth, all in UK. Since 1999, he has been with the School of Electronics and Computer Science, the University of Southampton, UK, where he holds the post of Professor in Intelligent Systems and Signal Processing. Dr Chen's research interests include adaptive signal processing, wireless communications, modeling and identification of nonlinear systems, neural network and machine learning, intelligent control system design, evolutionary computation methods and optimization. He has published over 600 research papers. Professor Chen has 19,400+ Web of Science citations with h-index 61 and 38,000+ Google Scholar citations with h-index 82. Dr. Chen is a Fellow of the United Kingdom Royal Academy of Engineering, a Fellow of Asia-Pacific Artificial Intelligence Association, a Fellow of IET, and a Life Fellow of IEEE. He is one of the original ISI highly cited researchers in engineering (March 2004).
\end{IEEEbiography}

\end{document}